\documentclass{ws-rv9x6}
\usepackage{subfigure}   
\usepackage{ws-rv-thm}   
\usepackage{ws-rv-van}   
\makeindex

\newcommand{\ket}[1]{\big| \,{#1}\, \big> }
\newcommand{\braket}[2]{\big< \,{#1}\, \big| \,{#2}\, \big> }
\newcommand{\matrixe}[3]{\big< \,{#1}\, \big| \,{#2}\, \big| \,{#3}\, \big> }
\newcommand{\op}[1]{\widehat{#1}}
\newcommand{\conj}[1]{{{#1}}^{\star}}
\newcommand{\hermit}[1]{{{#1}}^{\dag}}

\begin{document}

\chapter[Nuclear Clustering in Fermionic Molecular Dynamics]{Nuclear Clustering in Fermionic Molecular Dynamics}

\author[H. Feldmeier and T. Neff]{Hans Feldmeier$^{1,2}$ and Thomas Neff$^1$}

\address{$^1$GSI Helmholtzzentrum f{\"u}r Schwerionenforschung GmbH,\\
Planckstra{\ss}e 1, 64291 Darmstadt, Germany\\
$^2$Frankfurt Institute for Advanced Studies,\\
Max-von-Laue Stra{\ss}e, 60348 Frankfurt, Germany\\
h.feldmeier@gsi.de, t.neff@gsi.de}

\begin{abstract}
	Clustering plays an important role in the structure of nuclei, especially for light nuclei in the $p$-shell. In nuclear cluster models these degrees of freedom are introduced explicitly. In the Resonating Group Method or in the Generator Coordinate Method the clusters are built from individual nucleons interacting via an effective nucleon-nucleon interaction; the total wave function is antisymmetrized. Fermionic Molecular Dynamics (FMD) goes beyond pure cluster models. It is a microscopic many-body approach using a Gaussian wave packet basis that includes the harmonic oscillator shell model and Brink-type cluster model wave functions as special cases. Clustering is not imposed but appears dynamically in the calculations. The importance of clustering for the understanding of bound states, resonances and scattering states is illustrated with examples discussing the charge radii of the Neon isotopes, the $^3$He($\alpha$,$\gamma$)$^7$Be capture reaction and the cluster states in the $^{12}$C continuum.
\end{abstract}

\body


\section{General Considerations}\label{sec:general}

\subsection{Clusters, thresholds and the Resonating Group Method}\label{sec:clusters}

Let us first look at the textbook example for a nuclear cluster, the atomic nucleus of $^4$He or $\alpha$-particle. It is well bound with a binding energy of $E^{0^+}_1 = -28.30$~MeV. As the $^4$He nucleus is a rather compact system with a charge radius of about 1.67~fm all nucleons are so close that for each of the six possible pairs the nucleons are within the range of their nuclear interaction. When all pairs occupy the same $l$=0 state, i.e., the coordinate space part is completely symmetric under particle permutations, the Pauli principle allows three pairs of nucleons with $S$=1, $T$=0 and three pairs with $S$=0, $T$=1. On the other hand these two channels exhibit the strongest attraction in the nucleon-nucleon force. Therefore it is not surprising that the ground state is tightly bound and when calculated with a realistic interaction it consists to about 85\% of this simple configuration, which in shell model language corresponds to filling the lowest $0s$-shell with four nucleons. 

\begin{figure}[tb]
	\centering
	\includegraphics[width=0.4\textwidth]{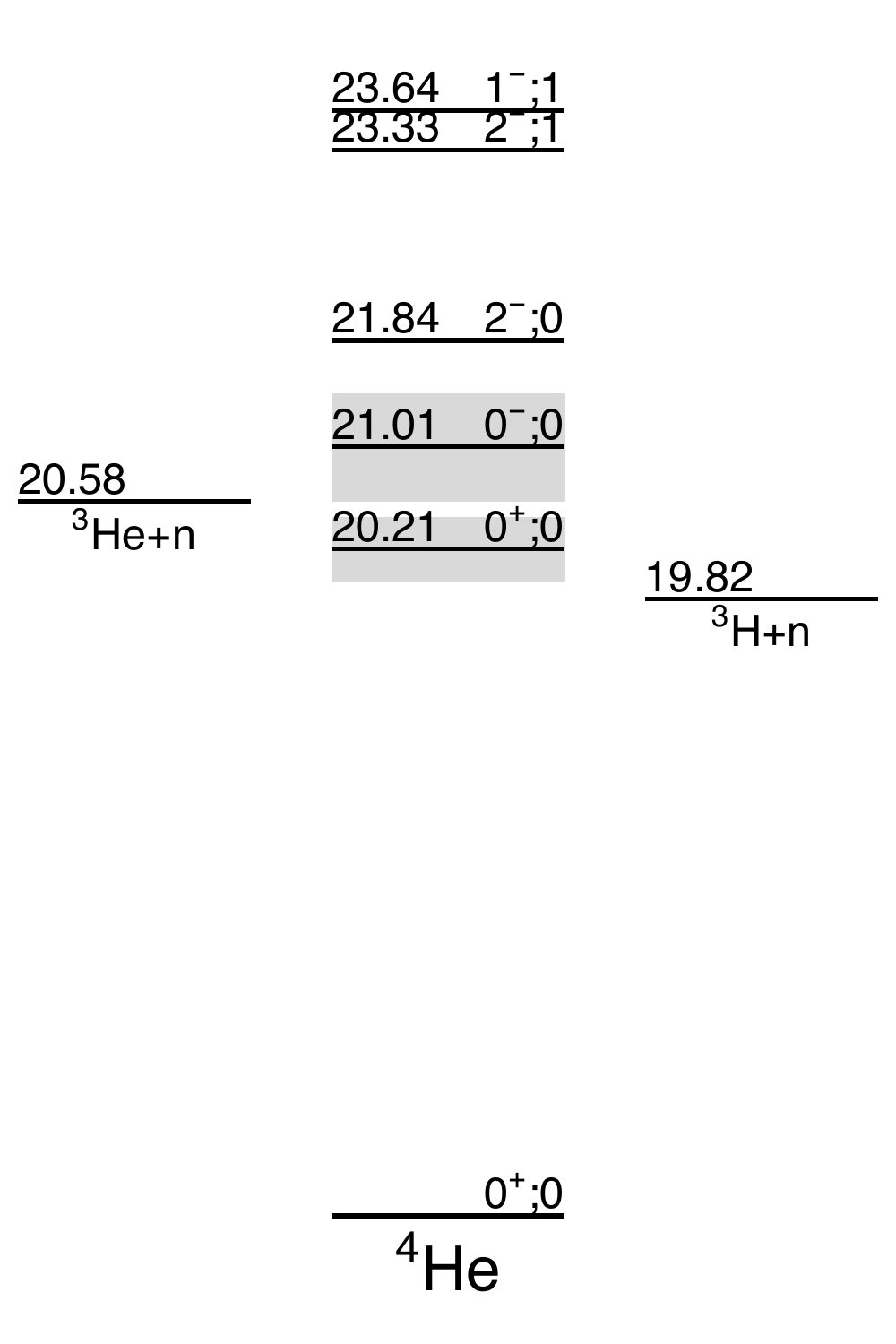}
	\caption{Energy spectrum of $^4$He with thresholds. Labels indicate $J^\pi;T$, energies in MeV (adapted from TUNL). For the resonances at 20.11 and 21.01~MeV the widths are indicated explicitly by shaded areas. The resonances above are rather broad and their widths overlap. The position of the ground state is not to scale.}
	\label{fig:4He}
\end{figure}

The other important aspect that makes the $\alpha$-particle an ideal cluster is that there is no further bound state in $^4$He. The first excited states appear above 20~MeV and are already resonances, $E^{0^+}_2$= 20.21~MeV, width  $\Gamma$ = 0.5~MeV and $E^{0^-}_1$= 21.01~MeV, $\Gamma$ = 0.84~MeV (see Fig.~\ref{fig:4He}). This implies that the $\alpha$-particle is not easily excited and difficult to polarize. Therefore one expects that the $\alpha$-cluster structure survives as a four-body correlation also in heavier nuclei. 

A special role is reserved for $\alpha$-nuclei that have an equal even number of protons and neutrons and can be thought of being composed of $A/4$ $\alpha$-clusters\cite{wigner37}. The idea of finding pronounced $\alpha$-cluster states at the corresponding thresholds is nicely summarized in the Ikeda diagram\cite{ikeda68}.

Generally one encounters noticeable cluster structures at energies in the vicinity of thresholds for breakup of a nucleus into various clusters,  {C~$\rightarrow$~A~+~B}. Let us first consider the case where the energy $E$ is just above the lowest threshold ($E>E(\mathrm{A})+E(\mathrm{B})-E(\mathrm{C})$) where the eigenstates of the $A$-body system C describe elastic scattering A + B $\rightarrow$ A + B.

At large distance $|\vec{r}_\mathrm{AB}|=|\vec{r}_\mathrm{A}-\vec{r}_\mathrm{B}|$ between A and B it is quite natural to describe the scattering system with RGM states 
\begin{equation}\label{eq:RGM1}
	\ket{\Phi_\mathrm{C}} = \int \!\mathrm{d}^3 r_\mathrm{AB} \,
  	\ket{\Phi_\mathrm{AB};\vec{r}_\mathrm{AB}} \: \varphi_\mathrm{AB}(\vec{r}_\mathrm{AB}) \: ,
\end{equation}
where $\varphi_\mathrm{AB}(\vec{r}_\mathrm{AB})$ represents the wave function characterizing the relative motion between the center-of-mass (c.m.) positions of clusters A and B. The wave function of the many-body RGM basis state $\ket{\Phi_\mathrm{AB};\vec{r}_\mathrm{AB}}$ labeled by the relative distance $\vec{r}_\mathrm{AB}=\vec{r}_\mathrm{A}-\vec{r}_\mathrm{B}$ between the c.m. of cluster A and B 
\begin{equation}\label{eq:RGM2}
	\braket{\xi_\mathrm{A},\xi_\mathrm{B},\vec{r}}{\Phi_\mathrm{AB};\vec{r}_\mathrm{AB}} =
	\op{\mathcal{A}} \: \big\{ \Phi_\mathrm{A}(\xi_\mathrm{A}) \Phi_\mathrm{B}(\xi_\mathrm{B}) \: \delta^3(\vec{r}-\vec{r}_\mathrm{AB}) \big\}
\end{equation}
is an antisymmetrized  product of a localized wave function for the relative motion times the intrinsic wave functions $\Phi_\mathrm{A}(\xi_\mathrm{A})$ and $\Phi_\mathrm{B}(\xi_\mathrm{B})$. The antisymmetrizer $\op{\mathcal{A}}$ projects on a state that is antisymmetric under all particle permutations, ($\op{\mathcal{A}}^2=\op{\mathcal{A}}, \hermit{\op{\mathcal{A}}}=\op{\mathcal{A}}$).
$\Phi_\mathrm{A}(\xi_\mathrm{A})$ and $\Phi_\mathrm{B}(\xi_\mathrm{B})$ denote respectively intrinsic eigenstates of A and B in terms of their sets of intrinsic variables
\begin{equation*}
	\xi_\mathrm{A} = \{ \vec{\xi}_i; i=1,\cdots,{A_\mathrm{A}-1}\} \cup \{ \sigma_i,\tau_i; i=1,\cdots,A_\mathrm{A}\}
\end{equation*}	
and
\begin{equation*}
	\xi_\mathrm{B} = \{ \vec{\xi}_j; j={A_\mathrm{A}+1},\cdots,{A_\mathrm{C}-1}\} \cup
                            \{\sigma_j,\tau_j; j={A_\mathrm{A}+1},\cdots,{A_\mathrm{C}} \} \: . 
\end{equation*}
$A_\mathrm{C}=A_\mathrm{B}+A_\mathrm{B}$ denote the mass numbers. There are altogether $A_\mathrm{C}-1$ independent coordinates, $A_\mathrm{A}-1$ in cluster A and $A_\mathrm{B}-1$ in cluster B plus the relative distance $\vec{r}$. Due to translational invariance there is no dependence on the total center-of-mass coordinate. Therefore in the following it will not be mentioned anymore, except when needed. Furthermore angular momentum coupling is not explicitly denoted here, but of course used in calculations.

As the antisymmetric RGM wave function $\braket{\xi_\mathrm{A},\xi_\mathrm{B},\vec{r}}{\Phi_\mathrm{AB};\vec{r}_\mathrm{AB}}$ is not a single Slater determinant it is not an easy task to perform the antisymmetrization for larger particle numbers. For example transposing a nucleon from A with one from B affects in general all intrinsic positions $\vec{\xi}_i$ in set $\xi_\mathrm{A}$ and all $\vec{\xi}_j$ in set $\xi_\mathrm{B}$ as they are defined in a translationally invariant way with respect to the distance $\vec{r}_\mathrm{A}-\vec{r}_\mathrm{B}$ of the clusters A and B.

The acronym RGM stands for Resonating Group Method\cite{wheeler37,tang78} introduced by J.A. Wheeler in 1937 where he proposed an antisymmetric many-body state consisting of groups of nucleons (clusters) that have a fixed internal structure but can move with respect to each other.

For energies above a threshold and at distances outside the range of the nuclear interaction between nucleons belonging to A or B the relative wave function $\varphi_\mathrm{AB}(\vec{r}_\mathrm{AB})$ has to match continuously the ingoing and outgoing Coulomb scattering states with the appropriate phase shift. Or in case of a narrow resonance described by a Gamow state it has to match an outgoing Coulomb wave with a complex relative momentum $k$, which encodes the width $\Gamma$. 

Being totally antisymmetric under particle permutations, the many-body basis state $\ket{\Phi_\mathrm{AB};\vec{r}_\mathrm{AB}}$ is a legitimate state at all $\vec{r}_\mathrm{AB}$, but for large overlaps of A and B at small $\vec{r}_\mathrm{AB}$, where the Pauli principle strikes, the antisymmetric part left over after the projection with $\op{\mathcal{A}}$ may be rather small. For a resonance, where the clusters A and B merge and stay together for a prolonged time, one expects in the interior the energy eigenstate to resemble more compact configurations typical for the compound nucleus C, while for intermediate distances one anticipates that deformed or polarized clusters A and B mix in. Finally the clusters separate at large distances as RGM states being in their ground states (or excited states if energy conservation allows). 

In $^4$He one encounters such a situation for
\begin{equation*}
	E(^3\mathrm{H})+E(\mathrm{p})-E(^4\mathrm{He})\ <\ E\ <\ E(^3\mathrm{He})+E(\mathrm{n})-E(^4\mathrm{He})
\end{equation*}
(see Fig.~\ref{fig:4He}). The 4-body system forms a resonance at $E(0^+_2)=20.21$~MeV, width $\Gamma=0.5$~MeV, very close to the threshold.
Fig.~\ref{fig:4He} also reveals that there is a second threshold at $E(^3\mathrm{He})+E(\mathrm{n})-E(^4\mathrm{He})=20.58~\mathrm{MeV}$ just above $E(0^+_2)=20.21$~MeV. This is quite an interesting situation as the channel $^3$He~+~n is not open yet, but very close in energy. Because the two thresholds are so close we expect mixing between the two cluster configurations such that at larger distances the RGM wave function looks like a coupled channel one, namely
\begin{align}
\nonumber
\ket{\Phi_{^4\mathrm{He}}} \overset{\text{large distance}}{\Longrightarrow}
 \ \ &\int\!\mathrm{d}^3 r \ 
\ket{\Phi_{^3\mathrm{H}} \Phi_\mathrm{p};\vec{r}} \: \varphi_{^3\mathrm{H,p}}(\vec{r})\\
+\ &\int\!\mathrm{d}^3 r \ 
\ket{\Phi_{^3\mathrm{He}}\Phi_\mathrm{n};\vec{r}} \: \varphi_{^3\mathrm{He,n}}(\vec{r})
\: .
\end{align}
For simplicity all spin and angular momentum couplings are not explicitly denoted. At large distances $\vec{r}$ the wave function has at the resonance energy the structure of a $^3$H~+~p scattering state with phase shifted Coulomb wave functions 
$\varphi_{^3\mathrm{H,p}}(\vec{r})$. Getting closer the second component $^3$He~+~n contributes with an exponentially decaying relative wave function 
$\varphi_{^3\mathrm{He,n}}(\vec{r}) \propto \exp(-k_{\,^3\mathrm{He,n}} r)$, where the relative momentum $k_{\,^3\mathrm{He,n}}=\sqrt{-2\mu \Delta}$ is given by the reduced mass $\mu$ and the energy difference $\Delta=E-E(^3\mathrm{He})-E(\mathrm{n})$ to the threshold in this channel. For the second resonance with quantum numbers $0_1^-$, where $E$ is above both thresholds, both channel wave functions have to be matched to phase shifted scattering solutions for negative parity.

To summarize: If the many-body system has an eigenenergy just below a breakup threshold the nuclear system already ``feels'' the continuum cluster structure in the outer part of the nucleus and the nucleons condense into the respective clusters. As we will show later in actual calculations these cluster structures can prevail to rather small distances. If the eigenenergy slips above the threshold one gets a resonance with an asymptotic wave function describing the two clusters in relative motion. The inner region of the $A$-body system may, but need not, resemble the cluster structure. One should also keep in mind that the antisymmetrizer $\op{\mathcal{A}}$ blurs the picture for strongly overlapping clusters.

\subsection{Role of Antisymmetrization}\label{sec:antisymmetrization}

\begin{figure}[tb]
	\centering
	\includegraphics[width=0.6\textwidth]{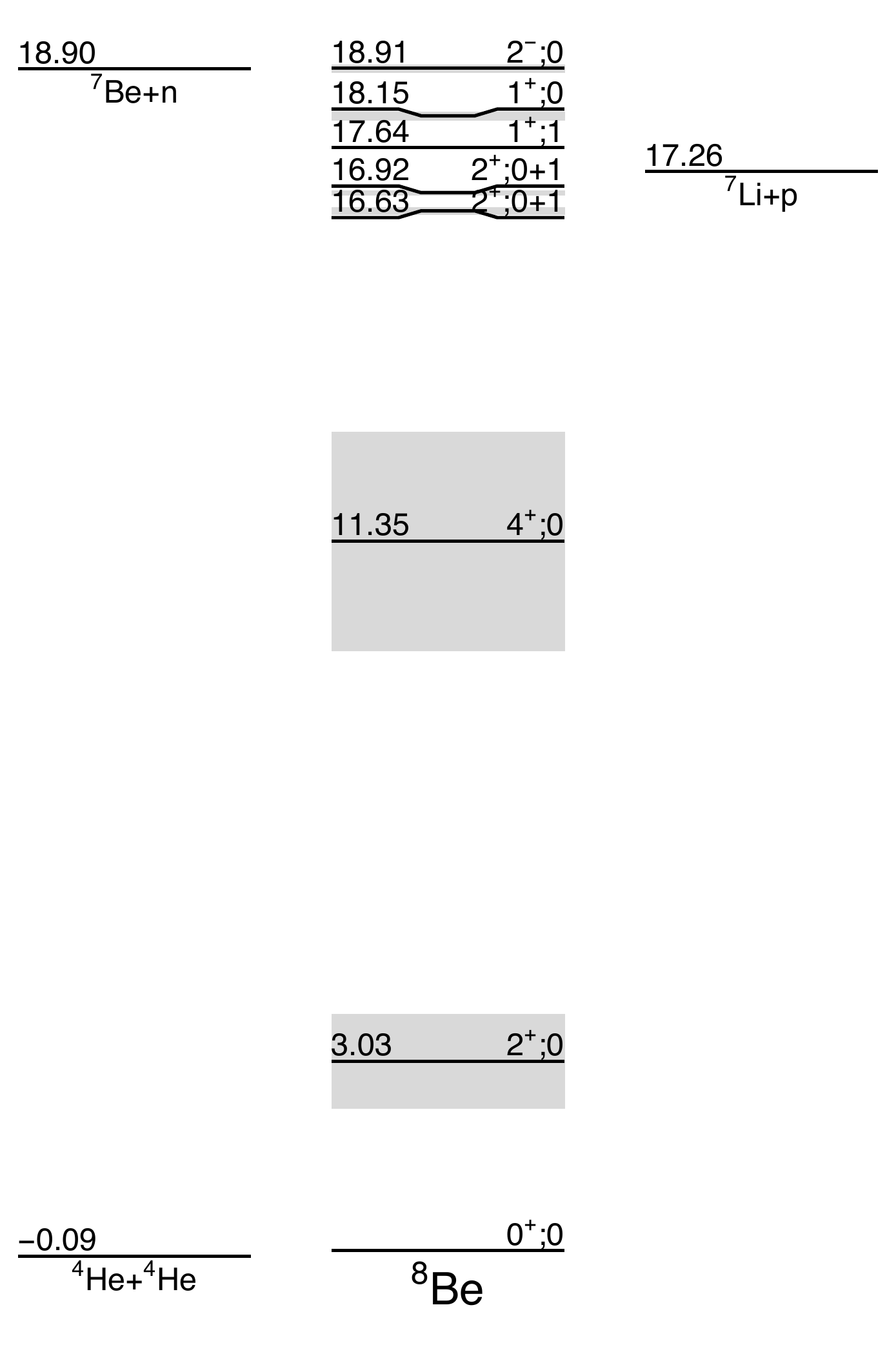}
	\caption{Energy spectrum of $^8$Be with thresholds. Labels indicate $J^\pi,T$, energies in MeV, widths are indicated by shaded areas (adapted from TUNL).}
	\label{fig:8Be}
\end{figure}

We discuss now the effect of the antisymmetrizer and then introduce the Generator Coordinate Method (GCM) for setting up a Slater determinant basis in $A$-body space that is numerically less painful than the RGM basis. We shall illustrate this with the example of $^8$Be, where there is only one threshold in the vicinity, so that coupled channels are not needed. From Fig.~\ref{fig:8Be} one sees that the ground state of $^8$Be is only 91.8~keV above the $^4$He~+~$^4$He threshold ($\alpha + \alpha$) and has the tiny width of $\Gamma = 5.6$~eV. The next threshold $^7$Li~+~p at 17.26~MeV will not influence the structure of the low lying resonances.

The structure of $^8$Be is a textbook example of clusterization. Having 4 neutrons and 4 protons the nuclear interaction chooses to first condense four nucleons into 2 strongly bound $\alpha$-clusters, respectively, thus gaining two times -28.30~MeV, and then arranging the relative motion of two $\alpha$-clusters such that the sum of kinetic energy, Coulomb repulsion and nuclear attraction is minimized. This implies that for a given inter-cluster potential the spatial localization of $\varphi_{\alpha\alpha}(\vec{r})$ should be washed out as much as possible  to keep the kinetic energy low. The localization in orientation $\hat{r}$ is completely removed by putting the relative motion into an $L=0$ state. 

The RGM wave function denoting explicitly the couplings takes the form
\begin{equation}\label{eq:RGM-8Be}
	\braket{\xi_\mathrm{A},\xi_\mathrm{B},\vec{r}}{\Phi_\mathrm{AB}^{(IL)J^\pi M}} = 
	\op{\mathcal{A}} \left\{ \left[\, \Bigl[\,\Phi_\mathrm{A}^{I_\mathrm{A}}(\xi_\mathrm{A}) \: \Phi_\mathrm{B}^{I_\mathrm{B}}(\xi_\mathrm{B}) \Bigr]^I Y^L(\hat{r}) \right]^{J^\pi M} \varphi_{\mathrm{AB}}^L(r) \right\} \: ,
\end{equation}
where $r=|\vec{r}|$  and $\hat{r}$ denotes the direction of $\vec{r}$. In the case of $^8$Be the wave functions $\Phi_\mathrm{A}^{I_\mathrm{A}}(\xi_\mathrm{A})$ and $\Phi_\mathrm{B}^{I_\mathrm{B}}(\xi_\mathrm{B})$ are $\alpha$-cluster states. Their intrinsic spins $I_\mathrm{A}=I_\mathrm{B}=0$ can only be coupled to the channel spin $I=0$ which in turn is coupled with the angular momentum $L$ of the relative motion to the total spin $J$. In this simple case $J=L$ and $\pi=(-1)^L$, where only even $L$ are allowed because A and B are identical bosons.

There are also the typical shell model states where 2 protons and 2 neutrons are in the $s$-shell and the remaining nucleons in the $p$-shell. Such shell model states are seen as excited states above 16~MeV (cf. Fig.~\ref{fig:8Be}) with a small decay width because their structure is quite different so that they cannot easily decay into $\alpha$-particles. Thus, first condensing four nucleons into $\alpha$-clusters and then arranging the two $\alpha$-clusters in an optimal way brings an energy advantage of 16~MeV over putting all 8 nucleons in a common mean field. 

It should be noted that there is a large overlap between shell model and cluster model wave functions for the ground state band members. The cluster model wave functions at small distances are equivalent to shell model wave functions because in the limit $r \rightarrow 0$ antisymmetrization projects onto harmonic oscillator states (see Sec.~\ref{sec:gcm} and Fig.~\ref{fig:BrinkR}). Therefore shell model configurations with $S=0$ and $T=0$ mix with the spatially extended cluster configurations where the energy is lowered by delocalization in the relative motion of the $\alpha$-clusters. At variance with that, the states above 16~MeV that have a spin-flip nature with $S=1$ or $T=1$ quantum numbers are shell model configurations that do not mix with the cluster configurations.

In the 3-$\alpha$ nucleus $^{12}$C the mean field or shell model configuration prevails, partly due to the contribution of the spin-orbit force, in the ground state but some reminiscence of cluster structure remains, as is also the case in $^{20}$Ne and $^{24}$Mg. In these nuclei $\alpha$-cluster state appear as excited states near the energy of $n$ $\alpha$ particles (see the discussion on $^{12}$C in Sec.~\ref{sec:c12}).

\begin{figure}[tb]
  \includegraphics[width=0.48\textwidth]{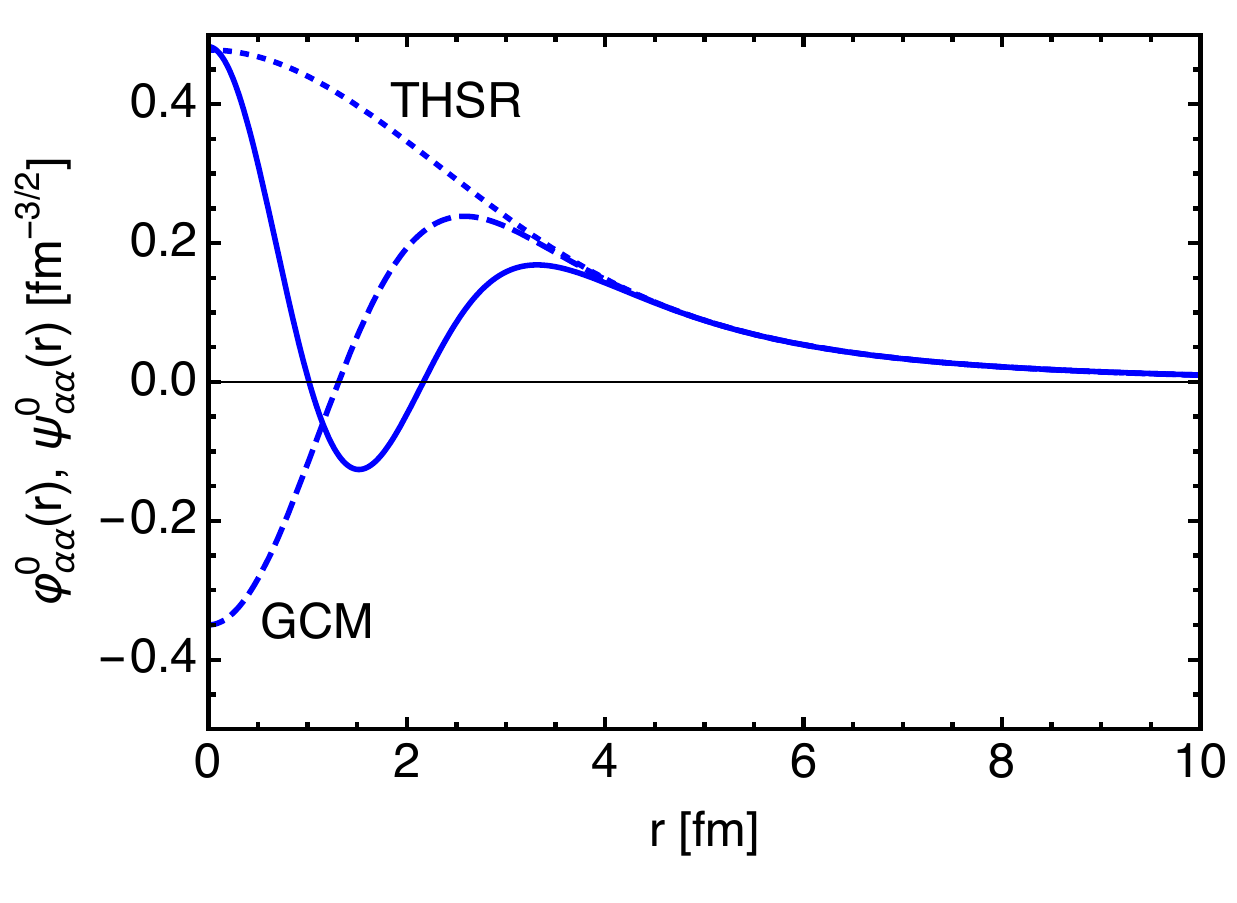}\hfil\includegraphics[width=0.48\textwidth]{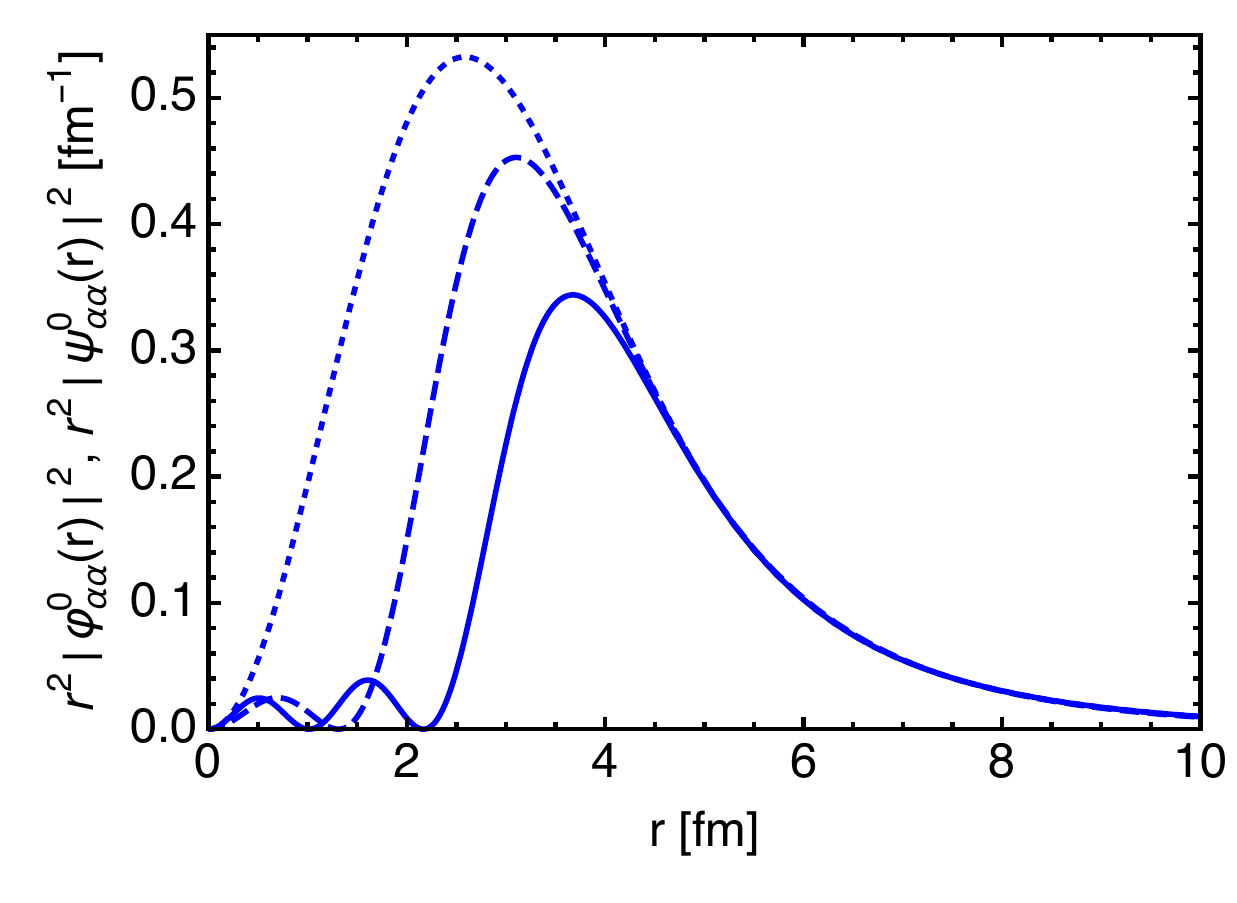}
	\caption{(Left) GCM (dashed) and THSR (dotted) relative wave functions $\varphi_{\alpha\alpha}^0(r)$ before antisymmetrization and $\psi_{\alpha\alpha}^0(r)$ after antisymmetrization (coinciding full lines). (Right) Corresponding densities multiplied with $r^2$.}          
	\label{fig:rgm}
\end{figure}

There is an important issue in the interpretation of $\varphi_{\alpha\alpha}^0(r)$. When the two clusters are well separated the antisymmetrizer $\op{\mathcal{A}}$ acts as a unit operator but when for smaller $r$ the clusters are overlapping $\op{\mathcal{A}}$ begins to project away the non-antisymmetric part of the $A$-body wave function resulting first in a reduced amplitude and then for $r \rightarrow 0$ to oscillations.

Fig.~\ref{fig:rgm} shows the radial part of the relative wave function $\varphi_{\alpha\alpha}(\vec{r}) = \varphi_{\alpha\alpha}^0(r)\,Y^0_0(\hat{r})$ for the ground state of $^8$Be in terms of a $^4$He~+~$^4$He cluster wave function (Eq.~\eqref{eq:RGM-8Be}) for two examples. For the intrinsic state of $^4$He the most simple ansatz is taken with all nucleons in $0s$ relative motion for an oscillator parameter reproducing the experimental r.m.s. radius.               

In the case of two $\alpha$-clusters the channel spin is $I=0$ because both clusters have spin $I_\mathrm{A}=I_\mathrm{B}=0$ and hence the relative orbital angular momentum $L=J$ so that the angular momentum projected norm kernel for the RGM basis states $\ket{\Phi_{\alpha\alpha}; \vec{r}}$ (cf. Eq.~\eqref{eq:RGM1}) projected on $L=J$ takes the form
\begin{equation}\label{eq:norm}
	n_J(r,r')= \int\!\mathrm{d}^2\Omega_r\; \mathrm{d}^2\Omega_r' \: 
	\sum_M Y^{L=J}_M(\Omega_r) \, \braket{\Phi_{\alpha\alpha};\vec{r}}{\Phi_{\alpha\alpha};\vec{r}'} \, \conj{Y^{L=J}_M(\Omega_r')} \: .
\end{equation}
The wave function
\begin{equation}\label{eq:psi}
	\psi_{\alpha\alpha}^{L=J}(r)=\int_0^\infty \! r'^2 \mathrm{d}r' \: n_J^{1/2}(r,r') \: \varphi_{\alpha\alpha}^{L=J}(r')
\end{equation}
takes the reduction due the Pauli principle into account and its absolute squared value represents the probability to find the $\alpha$-clusters at distance $r$.

The RGM wave functions before, $\varphi_{\alpha\alpha}^0(r)$, and after antisymmetrization, $\psi_{\alpha\alpha}^0(r)$, shown in Fig.~\ref{fig:rgm} solve the RGM equations with a nuclear Hamiltonian, that is they diagonalize the Hamiltonian in the many-body space spanned by the RGM basis states. One sees a vibrational zero-point motion of the $\alpha$-clusters with a maximum probability around $r=3.5$~fm and the almost exponential tail of the Whittaker function tunneling into the Coulomb barrier and reaching out to large distances. 

The effects of antisymmetrization are seen when comparing $\varphi_{\alpha\alpha}^0(r)$ and  $\psi_{\alpha\alpha}^0(r)$. They are quite different at small $r$ and coincide at large $r$.  The corresponding probabilities displayed on the right hand side of Fig.~\ref{fig:rgm} show that at $r < 3.5~\mathrm{fm}$ a strong suppression sets in and below $r \approx 2.5~\mathrm{fm}$ the probability drops practically to zero. The antisymmetrized $\psi_{\alpha\alpha}^0(r)$ develops nodes reflecting the existence of Pauli-forbidden states. 

At this point one should emphasize that $\varphi_{\alpha\alpha}^0(r)$ is not well defined for small $r$, which however does not matter because  physical observables are not affected by this. This is illustrated by the fact that two quite different relative wave functions $\varphi_{\alpha\alpha}^0(r)$, one given by the THSR\cite{funaki02} ansatz which has no nodes and one which is obtained using GCM basis states, as described in the next section, with one node are mapped by the antisymmetrization onto to the very same $\psi_{\alpha\alpha}^0(r)$ (full line in Fig.~\ref{fig:rgm}). Therefore any reference to properties of the choice of $\varphi_{\alpha\alpha}^0(r)$ in the region where the clusters have strong overlap has no relevance. 

For large $r$ beyond  $r \approx 5~\mathrm{fm}$ antisymmetrization is no longer essential and the norm kernel becomes local $n_J(r,r') \rightarrow N_J \delta(r-r')/(rr')$.


\subsection{Slater Determinants and Generator Coordinate Method}\label{sec:gcm}

The antisymmetrization which complicates the evaluation of Eq.~\eqref{eq:RGM-8Be} can be much more easily achieved if one uses Slater determinants to set up the many-body basis. These are often denoted as Generator Coordinate Method (GCM) states which are labeled and distinguished by generator coordinates. For example they can be generated by minimizing the energy of a Slater determinant in the Hartree-Fock method under a constraint that plays the role of the generate coordinate. Examples are various moments like mass quadrupole, octupole or charge dipole.

Another very simple example are Brink-type\cite{brink65} cluster states that are the analogue to the RGM basis states discussed above. Here one does not use intrinsic states for the clusters but Slater determinants of localized single-particle states. 
\begin{equation}\label{eq:Brink}
	\ket{\Psi_\mathrm{AB};\vec{R}_\mathrm{A},\vec{R}_\mathrm{B}}=
	\op{\mathcal{A}} \left\{ \ket{\Psi_\mathrm{A};\vec{R}_\mathrm{A}} \otimes \ket{\Psi_\mathrm{B};\vec{R}_\mathrm{B}} \right\}
\end{equation}
$\ket{\Psi_\mathrm{A};\vec{R}_\mathrm{A}}$ and $\ket{\Psi_\mathrm{B};\vec{R}_\mathrm{B}}$ are Slater determinants representing clusters A and B whose mean positions are shifted in space to $\vec{R}_\mathrm{A}$ and $\vec{R}_\mathrm{B}$, respectively. $\vec{R}_\mathrm{A}$ and $\vec{R}_\mathrm{B}$ are in this case the generator coordinates.

In general the wave function $\braket{\xi_\mathrm{A},\xi_\mathrm{B},\vec{r},\vec{r}_\mathit{cm}}{\Psi_\mathrm{AB};\vec{R}_\mathrm{A},\vec{R}_\mathrm{B}}$ does not separate into intrinsic times relative motion as was the case for the RGM basis. Also the total c.m. motion can in general not be factorized out. This is a disadvantage when treating scattering problems where boundary conditions have to be imposed on the relative motion, but for bound states where the clusters have usually strong overlap and antisymmetrization matters both RGM and GCM states span very similar many-body Hilbert spaces. The total c.m. motion inherent in Slater determinants can be removed by means of projection. Narrow resonances may be treated in good approximation as bound states, for which a representation in terms of RGM states at large distances are not needed.

For $^8$Be the most simple ansatz for the Brink state is a product of two $\alpha$-clusters where the spatial part of the single-particle states are harmonic oscillator $0s$-states, $\exp\big\{-(\vec{r}_i-\vec{R}_\mathrm{A,B})^2/(2a)\big\}$, displaced by the respective mean positions, $\vec{R}_\mathrm{A}$ and $\vec{R}_\mathrm{B}$. For that simple case the many-body wave function can be written as
\begin{multline}\label{eq:Brink-rel}
	\braket{\xi_\mathrm{A},\xi_\mathrm{B},\vec{r},\vec{r}_\mathit{cm}}{\Psi_{^8\mathrm{Be}};\vec{R}_\mathrm{A},\vec{R}_\mathrm{B}} = \\
  	\op{\mathcal{A}} \Bigl\{ \Phi_\alpha(\xi_\mathrm{A}) \: \Phi_\alpha(\xi_\mathrm{B}) \:
   		\varphi_\mathit{rel}(\vec{r}; \vec{R}_\mathrm{AB}^\mathit{rel}) \Bigr\} \: \varphi_\mathit{cm}(\vec{r}_\mathit{cm};\vec{R}_\mathrm{AB}^\mathit{cm})          
\end{multline}
with relative and center-of-mass wave functions
\begin{gather}\label{eq:Brink-rel-1}
	\varphi_\mathit{rel}(\vec{r};\vec{R}_\mathrm{AB}^\mathit{rel}) = \exp \biggl\{ -\frac{(\vec{r}-\vec{R}_\mathrm{AB}^\mathit{rel})^2}{2a/\mu_\mathrm{AB}}\biggr\} \:, \quad
	\vec{R}_\mathrm{AB}^\mathit{rel} = \vec{R}_\mathrm{A}-\vec{R}_\mathrm{B} \\
	\varphi_\mathit{cm}(\vec{r}_\mathit{cm};\vec{R}_\mathrm{AB}^\mathit{cm}) = \exp \biggl\{ -\frac{(\vec{r}_\mathit{cm}-	\vec{R}_\mathrm{AB}^\mathit{cm})^2}{2a/(A_\mathrm{A}+A_\mathrm{B})}\biggr\} \:, \quad
	\vec{R}_\mathrm{AB}^\mathit{cm} = \frac{A_\mathrm{A} \vec{R}_\mathrm{A}+A_\mathrm{B}\vec{R}_\mathrm{B}}{A_\mathrm{A}+A_\mathrm{B}}
\end{gather}
and the intrinsic wave functions
\begin{gather}\label{eq:Brink-int}
	\Phi_\alpha(\xi_\mathrm{A}) =
	\op{\mathcal{A}} \: \prod_{i=1}^4 \exp \biggl\{ -\frac{\vec{\xi}_i^2}{2a} \biggr\} \chi(\sigma_i,\tau_i)
          \quad \mathrm{with} \quad \vec{\xi}_i=\vec{r}_i-\frac{1}{A_\mathrm{A}} \sum_{l=1}^{A_\mathrm{A}} \vec{r}_l\\
	\Phi_\alpha(\xi_\mathrm{B}) =
	\op{\mathcal{A}} \: \prod_{j=5}^8 \exp \biggl\{ -\frac{\vec{\xi}_j^2}{2a} \biggr\} \chi(\sigma_j,\tau_j) 
          \quad \mathrm{with} \quad \vec{\xi}_j=\vec{r}_j-\frac{1}{A_\mathrm{B}} \sum_{l=A_\mathrm{A}+1}^{A_\mathrm{A}+A_\mathrm{B}} \vec{r}_l  \: ,   
\end{gather}
which depend on the intrinsic coordinates $\vec{\xi}_i$ that are measured from the respective c.m. coordinates. The spin, isospin orientations are denoted by $\chi(\sigma_i,\tau_i)$.

Comparing to the RGM basis wave function in Eq.~\eqref{eq:RGM2} one sees that the relative motion of the clusters is not sharply localized anymore but the $\delta$-function is smeared out to the Gaussian $\varphi_\mathit{rel}(\vec{r};\vec{R}_\mathrm{AB}^\mathit{rel})$ with mean position $\vec{R}_\mathrm{AB}^\mathit{rel}$ and a width given by the single-particle width parameter $a$ divided by the reduced mass number $\mu_\mathrm{AB}$ (=2 in the example). One should be aware that the relative distance 
\begin{equation*}
	\vec{r}=\frac{1}{A_\mathrm{A}} \sum_{i=1}^{A_\mathrm{A}} \vec{r}_i - \frac{1}{A_\mathrm{B}} \sum_{j=A_\mathrm{A}+1}^{A_\mathrm{A}+A_\mathrm{B}} \vec{r}_j
\end{equation*}
is not an independent variable anymore but depends on all single-particle positions.
The second difference is that there occurs a total center-of-mass wave function, which depends on $\vec{r}_\mathit{cm}=\frac{1}{A_\mathrm{A}+A_\mathrm{B}} \sum_{i=1}^{A_\mathrm{A}+A_\mathrm{B}} \vec{r}_i$ and thus on all single-particle positions. This was not present in the RGM case. Because of translational invariance results do not depend on the overall center-of-mass motion and hence it was not considered at all in RGM. 

In the general case the above factorization into intrinsic, relative, and c.m. motion does not hold. Slater determinants entangle in a spurious way intrinsic and c.m. motion, therefore in GCM this problem has to be addressed by means of projection on total c.m. momentum zero (see Sec.~\ref{sec:projection}).

\begin{figure}[tb]
  \centerline{
  \includegraphics[width=0.9\textwidth]{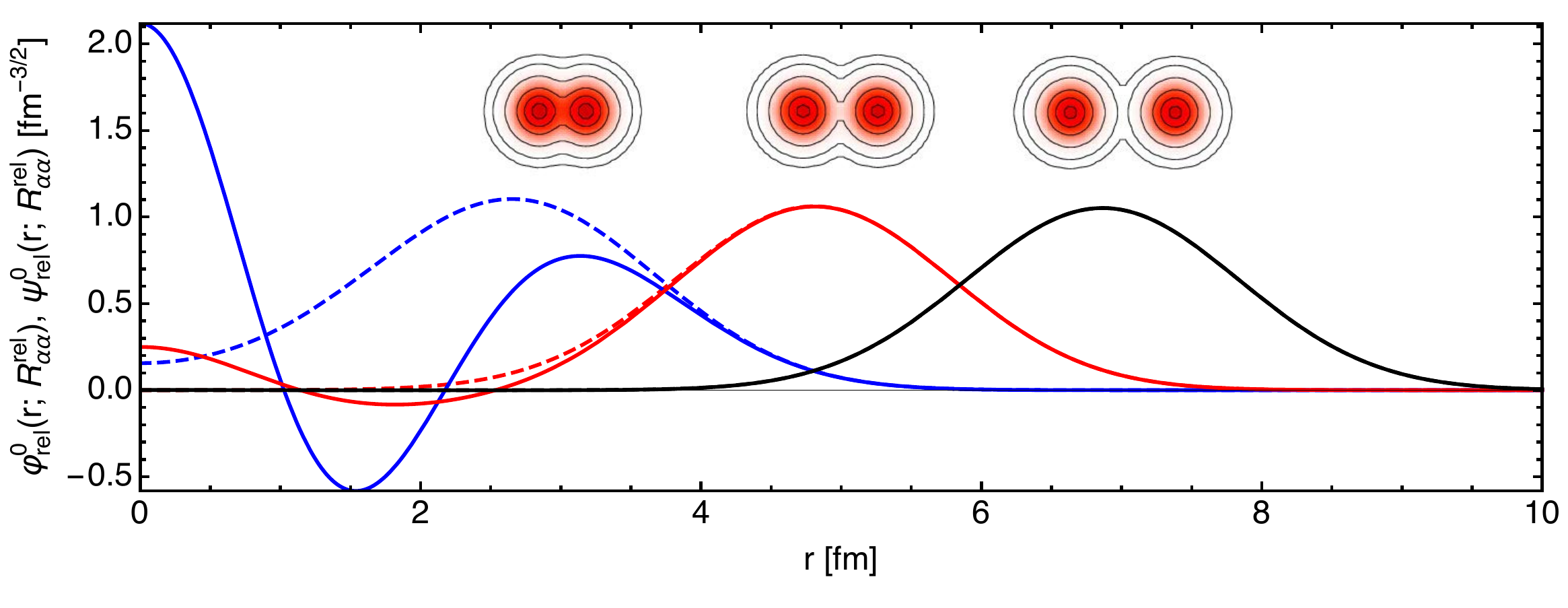}}
	\caption{Top: intrinsic one-body densities of Brink $\alpha-\alpha$ cluster states for mean distances $R_{\alpha\alpha}^	\mathit{rel}=3,5,7$~fm (blue, red, black). Bottom: corresponding relative wave functions $\varphi_\mathit{rel}^0(r; R_{\alpha\alpha}^\mathit{rel})$ before (dashed lines) and $\psi_\mathit{rel}^0(r; R_{\alpha\alpha}^
\mathit{rel})$ after antisymmetrization (full lines) as function of distance r.}                  
	\label{fig:BrinkR}
\end{figure}

Fig.~\ref{fig:BrinkR} displays in the upper part the intrinsic one-body density distribution that is obtained when aligning $\vec{R}_\mathrm{AB}^\mathit{rel}$ along the $x$-axis for $R_{\alpha\alpha}^\mathit{rel} \equiv |\vec{R}_\mathrm{AB}^\mathit{rel}| = 3,5,7$~fm. In the lower part the corresponding relative wave functions $\varphi_\mathit{rel}^0(r;R_{\alpha\alpha}^\mathit{rel})$ and the antisymmetrized  $\psi_\mathit{rel}^0(r;R_{\alpha\alpha}^\mathit{rel})$ are plotted as a function of $r$. They are obtained by projection of $\varphi_\mathit{rel}(\vec{r};\vec{R}_{\alpha\alpha}^\mathit{rel})$ and $\psi_\mathit{rel}(\vec{r};\vec{R}_{\alpha\alpha}^\mathit{rel})$ on $L=0$. The relative wave packet at $R_{\alpha\alpha}^\mathit{rel}=7$~fm is not influenced by antisymmetrization at all and the one at $R_{\alpha\alpha}^\mathit{rel}=5$~fm very little. But the one for $R_{\alpha\alpha}^\mathit{rel}=3$~fm, where the one-body density shows appreciable overlap of the $\alpha$-clusters, is completely changed into an oscillating wave resembling a $2s$ harmonic oscillator state with two nodes. This way the Pauli principle is restored. It is interesting to note the similarity at small $r$ to the antisymmetrized RGM state shown in Fig.~\ref{fig:rgm}. Apparently in the interior of the nucleus the antisymmetrization projects different wave functions onto the same state. The other information contained in this example is that for low energy a linear combination of GCM basis states taken at discrete values of $\vec{R}_\mathrm{AB}^\mathit{rel}$ will suffice to generate a smooth relative wave function that is completely equivalent to the RGM wave function.

To summarize: RGM states are easily matched to boundary conditions but increasingly difficult to antisymmetrize with growing particle number. GCM states are easy to antisymmetrize but are more difficult to match to scattering solutions. The GCM basis and the RGM basis are equivalent. For bound states and narrow resonances, treated in bound state approximation, the GCM basis is preferable because it is numerically easier, more flexible and can include also non-cluster states.

\section{Fermionic Molecular Dynamics}\label{sec:fmd}

Cluster models can only be understood as an approximation to a full description of the nucleus. The wave functions of the clusters are typically restricted to simple harmonic oscillator shell model functions. The description by purely cluster degrees of freedom is especially questionable at short distances where the clusters strongly overlap and where one expects a polarization of the clusters. Even in the case of $^8$Be the $\alpha$-cluster picture has to be modified. At short distances shell model configurations with a preferred occupation of $p_{3/2}$ over $p_{1/2}$ orbits due to the spin-orbit force should be important. In a pure cluster model such polarization and admixture effects are neglected and at best may be absorbed in the effective interaction. As these effects will strongly depend on the nucleus and even on the state under consideration it is not surprising that the effective interactions used in cluster models are not universal and are typically tuned for each nucleus. These issues become especially challenging for nuclei which are not $\alpha$-nuclei.

Fermionic Molecular Dynamics (FMD)\cite{fmd90,fmd00,fmd04b,fmd04,fmd08} is a many-body approach that can be understood as an extension of traditional cluster models. The effective degrees of freedom are here not restricted to the relative motion of a given set of clusters but the individual nucleons are considered as the degrees of freedom. The single-particle states are described by Gaussian wave-packets localized in phase space. In FMD clusters should emerge automatically as effective degrees of freedom if the system prefers a breakup into clusters. In this way FMD allows for a consistent description of clustering in all nuclei. An important ingredient is also the use of a realistic effective interaction that is not tuned to particular systems.

\subsection{Intrinsic States}\label{sec:intrinsic}

FMD is a fully microscopic many-body approach that uses Slater determinants as intrinsic many-body basis states
\begin{equation}
  \label{eq:fmdslaterdet}
  \ket{Q} = \op{\mathcal{A}} \bigl\{ \ket{q_1} \otimes \ldots \otimes \ket{q_A} \bigr\} \: .
\end{equation}
The many-body state $\ket{Q}$ is given by $A$ single-particle states $\ket{q_k}$  
\begin{equation}
  \ket{q_k} = \sum_{i=1}^N \ket{a_{ki} \vec{b}_{ki}} \otimes
  \ket{\chi^\uparrow_{ki},\chi^\downarrow_{ki}} \otimes \ket{\xi_k} \: c_{ki} \: ,
\end{equation}
where $N$ is typically one or two. The spatial part of the single-particle wave function is given by Gaussian wave packets
\begin{equation}
  \label{eq:gaussian}
  \braket{\vec{x}}{a,\vec{b}}= \exp \biggl\{ -\frac{(\vec{x} -\vec{b})^2}{2 a} \biggr\}
\end{equation}
and a spin-part given by a spinor $\ket{\chi^\uparrow,\chi^\downarrow}$. A single-particle state describes either a proton or a neutron, given by $\ket{\xi}$. The parameters of the wave packet are the width $a$ and the complex vector $\vec{b}$ that encodes the mean position and the mean momentum of the wave packet. 

An important point for understanding the versatility of the FMD basis is that it contains both harmonic oscillator shell model and Brink-type cluster wave functions as special cases. Harmonic oscillator single-particle states can be obtained by linear combinations of Gaussians in the limit of zero displacement. These linear combinations are automatically included in the Slater determinant, as the Slater determinant is not changed under linear transformations of the Gaussian single-particle states. Brink-type cluster wave functions are generated by localizing clusters of nucleons in phase space.

Gaussian wave packets have also other favorable properties. Translations, boosts and rotations of Gaussian wave packets are again Gaussian wave packets (with transformed parameters). These properties of the Gaussian wave packets are inherited by the Slater determinants and it is straightforward to implement these transformations. Furthermore matrix elements of operators that are polynomial in coordinate or momentum space or have a Gaussian radial dependence can be evaluated analytically. A technical disadvantage is the fact that the Gaussian single-particle states are non-orthogonal which requires the calculation of overlap matrices.

\subsection{Restoration of Symmetries}\label{sec:projection}

The Slater determinants $\ket{Q}$ in general do not posses the symmetries of the Hamiltonian with respect to translations, rotations and reflections. These symmetries can be restored however by projection on parity, angular momentum, and total linear momentum:
\begin{equation}
  \ket{Q; J^\pi MK; \vec{P}=0} = 
  \op{P}^\pi \op{P}^{J}_{MK} \op{P}^{\vec{P}=0} \ket{Q} \: ,
\end{equation}
with the projection operators for parity
\begin{equation}
  \op{P}^\pi = \frac{1}{2} (1 + \pi \op{\Pi}) \: , 
\end{equation}
three dimensional angular momentum projection 
\begin{equation}
  \op{P}^{J}_{MK} = \frac{2J+1}{8\pi^2} \int d\Omega \; {D^J_{MK}}^\star(\Omega) \: \op{R}(\Omega) 
\end{equation}
with the rotation operator $\op{R}(\Omega)$ and the Wigner $D$-functions, and the projection on total linear momentum
\begin{equation}
  \op{P}^{\vec{P}} = \frac{1}{(2\pi)^3} \int d^3X \: \exp\{-i
  (\op{\vec{P}}-\vec{P})\cdot{\vec{X}} \} \: .
\end{equation}

As the Hamiltonian commutes with the symmetry operations it is sufficient to project once and calculate projected matrix elements for intrinsic basis states $\ket{Q^{(a)}}$ and $\ket{Q^{(b)}}$ for the Hamiltonian
\begin{equation}
  \mathsf{H}^{J^\pi}_{aK,bK'} = 
 	\matrixe{Q^{(a)}}{(\op{H} - \op{T}_\mathrm{cm}) 
    \op{P}^\pi \op{P}^J_{KK'} \op{P}^{\vec{P} = 0}}{Q^{(b)}} 
\end{equation}
and the overlap
\begin{equation}
  \mathsf{N}^{J^\pi}_{aK,bK'} = 
    \matrixe{Q^{(a)}}{\op{P}^\pi  \op{P}^J_{KK'} \op{P}^{\vec{P} = 0}}{Q^{(b)}}  \: .
\end{equation}
Multiconfiguration mixing eigenstates of the Hamiltonian are finally obtained by solving the generalized eigenvalue problem in a set of basis states $\left \{ \ket{Q^{(a)}} \right\}$
\begin{equation}
  \sum_{bK'} \mathsf{H}^{J^\pi}_{aK,bK'} \Psi^{J^\pi \alpha}_{bK'} =
  E^{J^\pi \alpha} \sum_{bK'} \mathsf{N}^{J^\pi}_{aK,bK'} \Psi^{J^\pi \alpha}_{bK'} \: .
\end{equation}

\subsection{Determination of Basis States}

The non-orthogonal and continuous nature of the wave packet basis does not lead to a unique choice of basis states. 
The basic idea in FMD is to generate a small number of optimized basis states by a variational procedure. This can be done on different levels of sophistication and effort. The simplest and numerically cheapest approach is a variation on the level of a single Slater determinant as in a mean-field calculation. Here one minimizes the energy of the intrinsic many-body state $\ket{Q}$ with respect to the parameters of the single-particle states
\begin{equation}
  \min_{\{q_\nu\}}
  \frac{\matrixe{Q}{\op{H}-\op{T}_\mathrm{cm}}{Q}}{\braket{Q}{Q}} \: .
\end{equation}
Symmetries can then be restored by projection, defining the \emph{projection after variation} (PAV) procedure. This is numerically cheap as the projection only has to be performed for the final basis state and in general works reasonably well for the description of deformed states that form a rotational band. However even in such cases correlation energies are underestimated.

An improved description is obtained by \emph{variation after projection} (VAP). Here the energy of the projected state
\begin{equation}
  \min_{\{q_\nu, C_K \}}
  \frac{\sum_{K,K'} \conj{C_K} \matrixe{Q}{(\op{H}-\op{T}_\mathrm{cm}) \op{P}^\pi \op{P}^J_{KK'} \op{P}^{\vec{P}=0}}{Q} C_{K'}}{\sum_{K,K'} \conj{C_K} \matrixe{Q}{\op{P}^\pi \op{P}^J_{KK'} \op{P}^{\vec{P}=0}}{Q} C_{K'}} 
\end{equation}
is minimized with respect to the single-particle parameters $q_\nu$ and the $K$-mixing parameters $C_K$. The VAP procedure is essential for states that have a different intrinsic structure than the mean-field state. This can be related to clustering but also to single-particle excitations. The VAP procedure will be performed for all quantum numbers of interest which provides a set of intrinsic basis states $\left\{ \ket{Q^{(a)}} \right\}$. The diagonalization of the Hamiltonian in the basis states obtained by projecting this set on the desired symmetries will further improve the description of the individual eigenstates due to the possible admixture of other configurations with the same quantum numbers.

A further improvement can be obtained by adding basis states obtained by variation under constraints. Constraints are related to collective degrees of freedom that are treated as generator coordinates, like the radius or quadrupole deformation. Take the case of a nucleus that has a prolate and an oblate minimum. VAP will provide the configuration with the lowest energy. The other minimum can however be found by minimizing the energy under constraints on the quadrupole deformation. Constraints are also very useful to obtain basis states for loosely bound systems of cluster or halo nature. The VAP minimum corresponds typically to a rather compact configuration. To include extended configurations necessary for the description of the tail of the wave function the variation can be performed under constraints on the radius.

\subsection{Realistic effective interaction}

An important ingredient in the FMD approach is the use of a realistic effective interaction. Realistic interactions, as used in \emph{ab initio} approaches, reproduce the nucleon nucleon scattering data up to the pion production threshold and the deuteron properties. However this does not allow a unique determination of the interaction. Whereas the long-range properties of the interaction are determined by pion dynamics the short-range behavior of the interaction is not fully constrained by the scattering data. In the Argonne~$v_{18}$ interaction \cite{wiringa95} the short-range behavior is modeled phenomenologically as a local interaction, whereas in interactions derived in chiral EFT \cite{entem03,epelbaum05} the short-range behavior is described by non-locally regularized contact terms. Despite these differences all realistic interactions show a strong short-range repulsion and a strong tensor force that make the application of realistic interactions in FMD or other many-body approaches difficult or impossible.

As in many \emph{ab initio} approaches a unitary transformation that decouples the low- and high-momentum components is used to derive an effective low-momentum interaction. In FMD we use the Unitary Correlation Operator Method (UCOM) that explicitly introduces central and tensor correlations. The UCOM approach has the advantage that it provides an explicit operator representation of the transformed interaction \cite{ucom03,ucom10,weber14}. The unitary transformation is performed in two-body approximation. Therefore two-body properties like nucleon nucleon phase shifts are conserved, but differences will appear in many-body system due to omitted three- and higher body terms in the transformed interaction. For light nuclei the two-body UCOM interaction works remarkably well as there appears to be a partial cancellation between the omitted three-body terms and genuine three-body forces that are not included. However the UCOM interaction shows deficiencies with respect to spin-orbit properties and for the saturation of heavier nuclei.

\section{Clustering in Bound States}\label{sec:neons}

\begin{figure}[b]
  \includegraphics[width=0.49\textwidth]{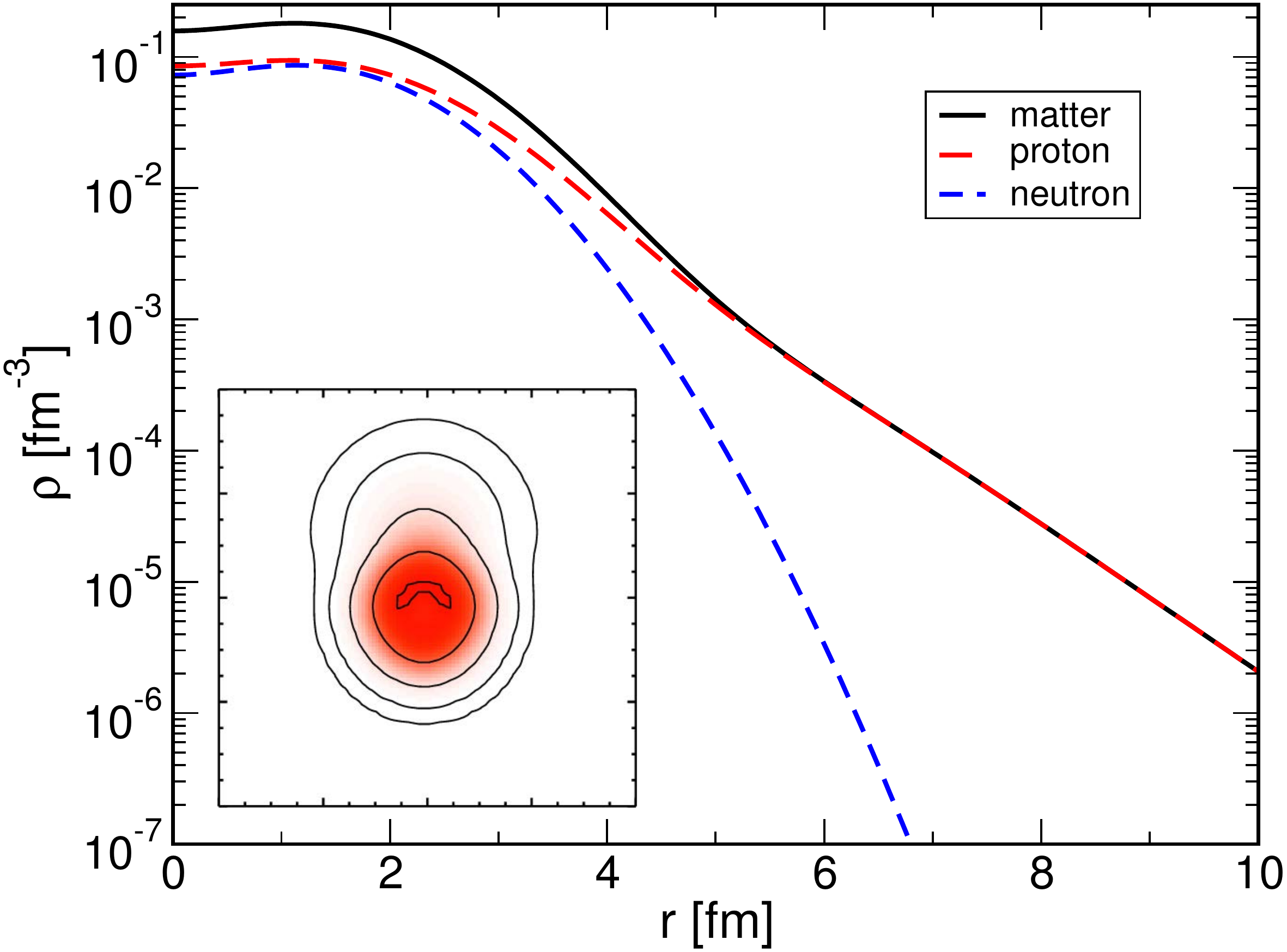}\hfil\includegraphics[width=0.48\textwidth]{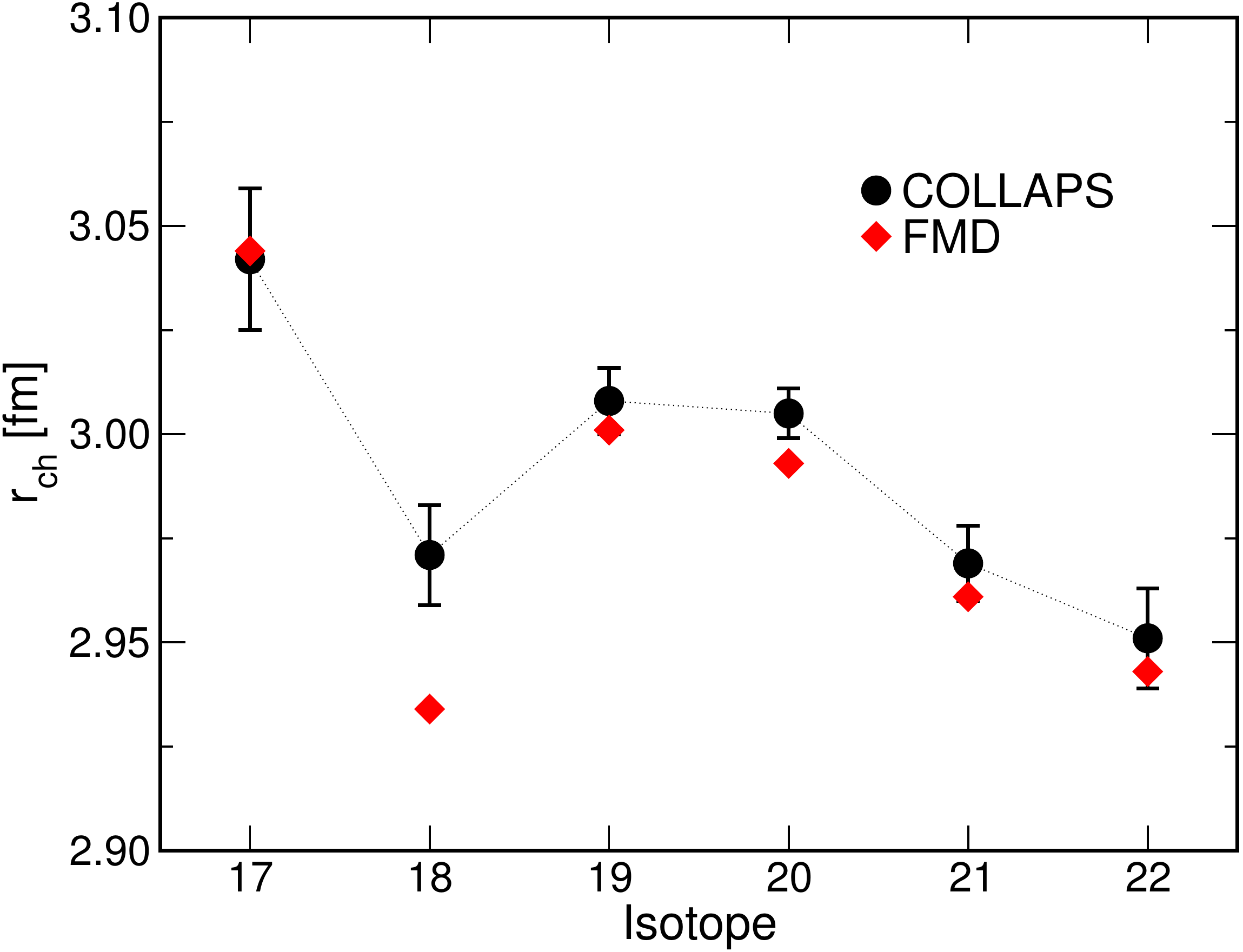}
  \caption{(Left) $^{17}$Ne proton and neutron point density distributions. The inset shows the intrinsic density of the dominant FMD configuration. (Right) Charge radii for $^{17-22}$Ne calculated in FMD compared to experimental results from COLLAPS.}
  \label{fig:neons} 
\end{figure}

A nice example for the importance of clustering is given by the Neon isotopes. Our investigation \cite{geithner08} was triggered by experimental measurements of the charge radii for $^{17-22}$Ne by the COLLAPS collaboration. Of particular interest here was the structure of $^{17}$Ne which has a very small two-proton separation energy of 0.93~MeV. In three-body calculations assuming an $^{15}$O core a large $s^2$ component and therefore halo-like structures were predicted, whereas shell model calculations based on Coulomb displacement energies and magnetic moments predicted only a small $s^2$ component. 

In the FMD calculations we performed variation after parity projection calculations with an additional constraint on the radius as generator coordinate. For $^{17}$Ne we found two minima that correspond essentially to an $^{15}$O core with two protons in a mixed $s^2$/$d^2$ configuration. These configurations strongly mix and in the multiconfiguration mixing calculation we find that the $s^2$ admixture in the $^{17}$Ne wave function is about 42\%. Due to the small two-proton separation energy and the missing centrifugal barrier in the $s$-orbit the proton distribution is rather extended which explains the large charge radius of $^{17}$Ne. The left part of Fig.~\ref{fig:neons} shows the extended tail of the proton distributions compared to the compact neutron distribution. The intrinsic density of the dominant FMD configuration further illustrates the correlations of the two valence protons. They like to form a pair on one side of the $^{15}$O core. This asymmetric shape can only be obtained by a linear combination of $s^2$ and $d^2$ configurations. Going from $^{17}$Ne to $^{18}$Ne we find a similar picture with an $^{16}$O core and two protons in $s^2$ and $d^2$ configurations. However the admixture of the $s^2$ component is only 15\% explaining the much smaller charge radius of $^{18}$Ne.

\begin{figure}[b]
	\includegraphics[width=0.3\textwidth]{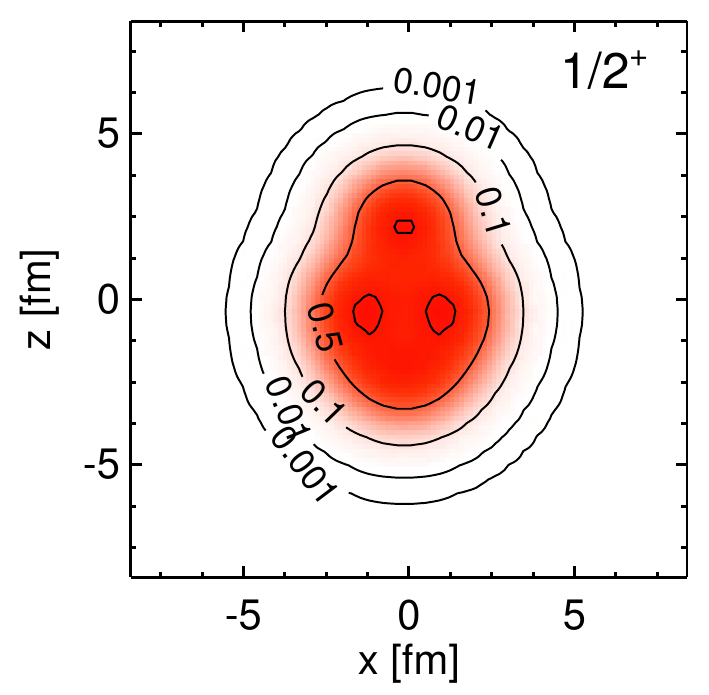}\includegraphics[width=0.3\textwidth]{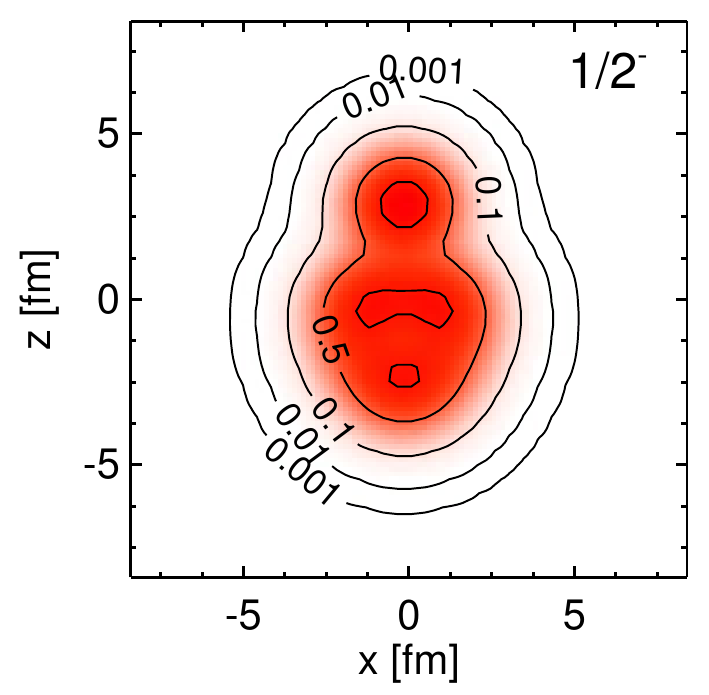}\hfil\includegraphics[width=0.3\textwidth]{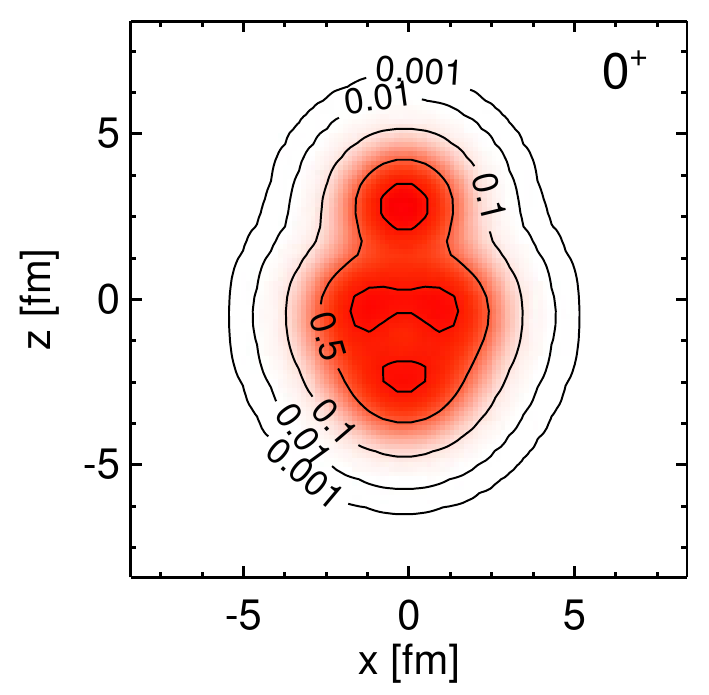}
	\caption{Intrinsic densities of dominant configurations for the $1/2^+$ ground state of $^{19}$Ne (left), the $1/2^-$ state in $^{19}$Ne (middle) and the $0^+$ ground state of $^{20}$Ne.}
	\label{fig:ne19-ne20}
\end{figure}

Interestingly the charge radii of $^{19}$Ne and $^{20}$Ne are again much larger. Another surprising observation is that experimentally the $1/2^-$ state in $^{19}$Ne is almost degenerate with the natural parity $1/2^+$ state. These observations, together with the rather low thresholds, already hint at the importance of $^4$He and $^3$He cluster configurations. In the FMD calculations we therefore not only included configurations optimized for positive parity but also configurations optimized for negative parity as shown in Fig.~\ref{fig:ne19-ne20}. Indeed we find in the multiconfiguration mixing calculations that in the $^{19}$Ne ground state extended configurations with both $^4$He+$^{15}$O and $^3$He+$^{16}$O cluster configurations contribute and are responsible for the large charge radius. In $^{20}$Ne the $^4$He+$^{16}$O threshold is only 4.5~MeV above the ground state and we find a significant admixture of extended cluster configurations. In the heavier Neon isotopes the admixture of $^4$He cluster configurations still exists but is much smaller than in $^{19}$Ne and $^{20}$Ne and correspondingly the charge radii are smaller.

Similar studies have been performed for the Lithium \cite{noertershaeuser11} and  Beryllium isotopes \cite{zakova10,krieger12,terashima14}. Also here the interplay between single-particle, especially the formation of halos, and cluster structure is reflected in the evolution of the charge radii.

\section{Clusters and Nuclear Reactions}\label{sec:capture}

Obviously clusters play an essential role in the description of nuclear reactions as the initial (and possibly final) states are given by asymptotically free clusters. Within the FMD approach the first application was the $^3$He($\alpha$,$\gamma$)$^7$Be radiative capture reaction \cite{neff11}. To calculate the reaction cross section one needs the initial scattering states, the final bound states and the electromagnetic transition matrix elements. To perform the calculation the model space is divided into two regions. In the external region bound and scattering states are described by $^3$He and $^4$He clusters in their FMD ground states. In the interaction region additional FMD configurations obtained by variation after parity and angular momentum projection on $1/2^+$, $3/2^+$, $5/2^+$, $1/2^-$, $3/2^-$, $5/2^-$ and $7/2^-$ are included. The radius is used in addition as a generator coordinate. These FMD configurations go beyond the simple cluster model picture and are essential for obtaining polarized cluster configurations. 

The bound and scattering eigenstates are determined by diagonalization of the Hamiltonian in the full space consisting of the internal and external region with the corresponding boundary conditions. In the external region only the Coulomb interaction between the clusters has to be considered and the system can be described as point-like clusters with the relative motion given by Coulomb wave functions depending on scattering energy and phase shift or by Whittaker functions depending on the bound state energies. The matching between the microscopic world of $A$-nucleon wave functions and the world of point-like clusters is done using the microscopic $R$-matrix method of the Brussels group \cite{descouvemont10}. Technically this requires a rewriting of the FMD Brink-type wave functions as RGM wave functions as discussed in Sec.~\ref{sec:gcm}. As the width parameters in the $^4$He and $^3$He cluster wave functions are different it is essential to perform the projection on total linear momentum. Only then the total wave function factorizes into the intrinsic wave functions of the clusters, the wave function describing their relative motion and the total (plane wave) motion of the center of mass. 

The properties of the bound states agree well with experimental data, but only if the polarized configurations are included in the model space. With the frozen cluster configurations alone, $^7$Be is bound only by 200~keV, in the full model space the $3/2^-$ state is bound by 1.49~MeV and the $1/2^-$ state by 1.31~MeV with respect to the cluster threshold. The calculated charge radius of 2.67~fm is also in good agreement with the experimental value of 2.647(17)~fm. For the phase shifts of the $S$- and $D$-waves we also find good agreement with experimental data. Again it is important to include the polarized configurations.  

\begin{figure}[tb]
	\includegraphics[width=0.48\textwidth]{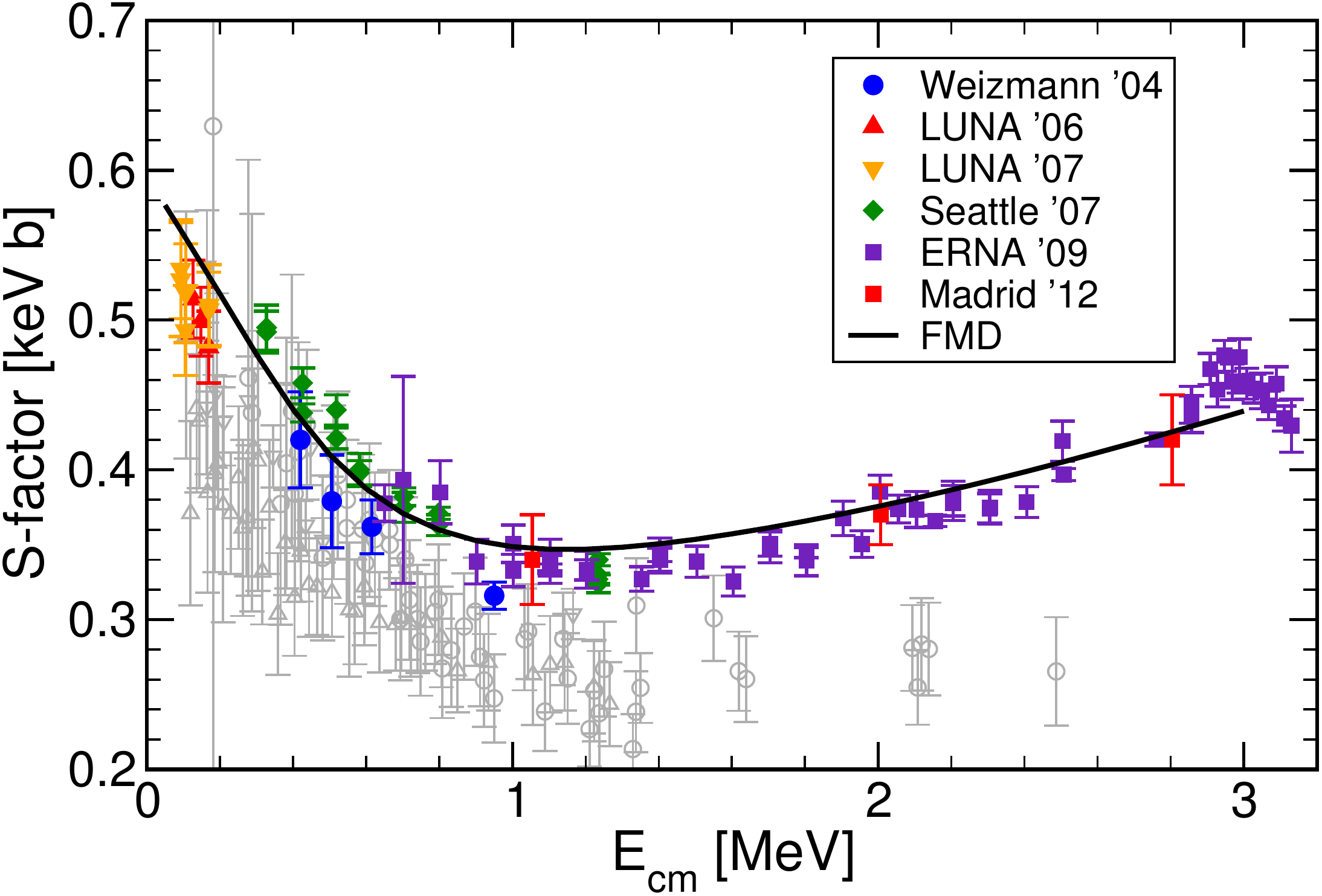}\hfil\includegraphics[width=0.49\textwidth]{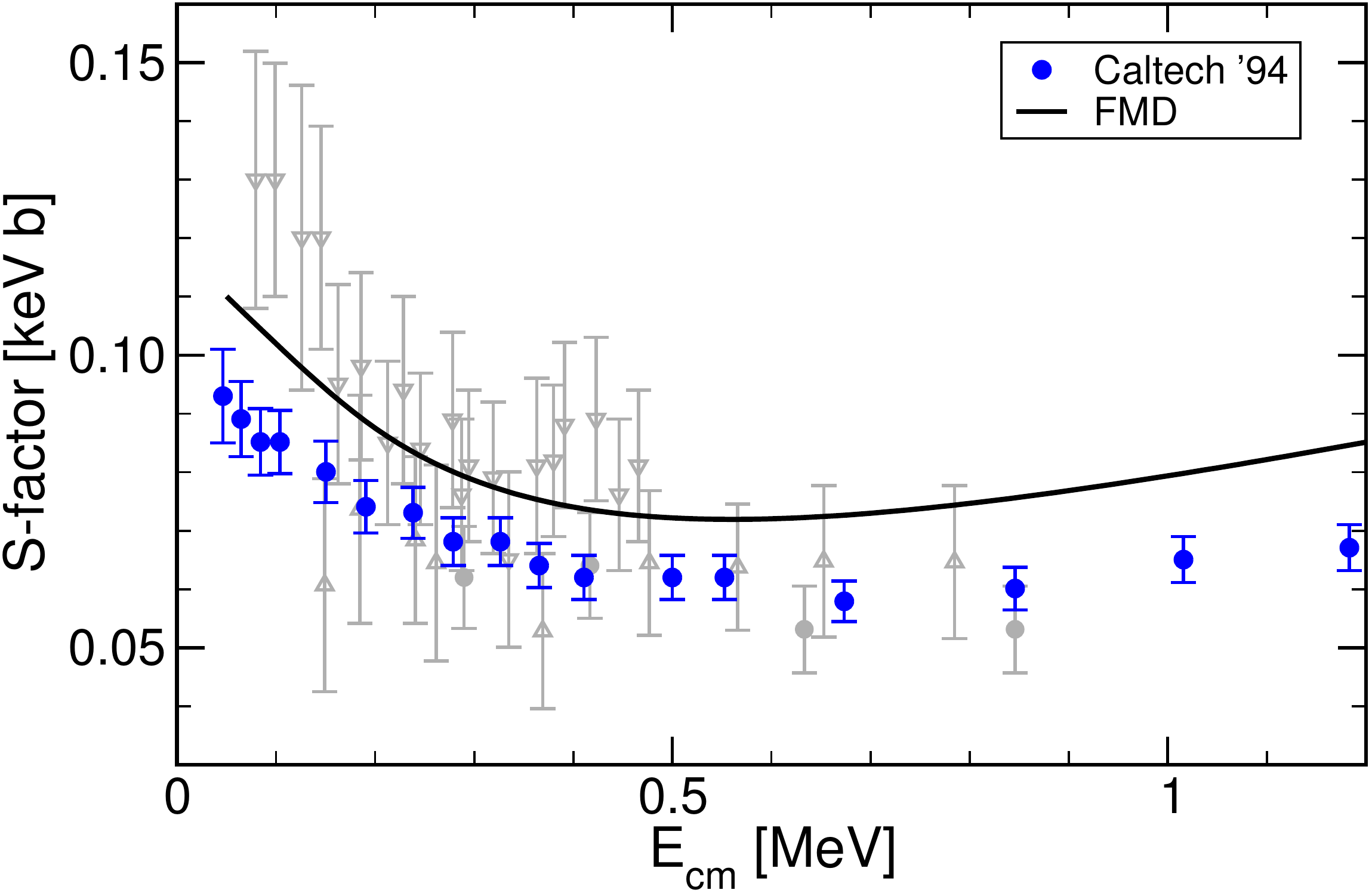}
	\caption{Astrophysical S-factors for the $^3$He($\alpha$,$\gamma$)$^7$Be and $^3$H($\alpha$,$\gamma$)$^7$Li capture reactions. Light gray data points indicate older experimental data.}
	\label{fig:capture}
\end{figure}

The calculated cross section for $^3$He($\alpha$,$\gamma$)$^7$Be, shown in the left part of Fig.~\ref{fig:capture} in form of the astrophysical $S$-factor agrees remarkably well with recent experimental data, both with respect to the energy dependence and the absolute normalization. In case of the mirror reaction $^3$H($\alpha$,$\gamma$)$^7$Li, shown in the right part of Fig.~\ref{fig:capture} the calculated energy dependence agrees perfectly with the data but the normalization is too large by about 15\%.

\section{Clustering in Resonance States}\label{sec:c12}

The structure of $^{12}$C above the three-$\alpha$ threshold poses a challenge for nuclear theory. The ground state band can be well described in the no-core shell model using a harmonic oscillator single-particle basis. However, many of the states in the continuum have a well developed cluster structure, and these states are completely missing in the no-core shell model \cite{maris14}. Microscopic $\alpha$-cluster models have been able to reproduce many properties of these continuum states \cite{kamimura81}. On the other hand, the cluster model is an idealization. Experimentally Gamow-Teller as well as $M1$ and $E1$ transitions into continuum states can be observed. Within a cluster model such transitions are forbidden, indicating that for a full picture both shell and cluster structure have to be included in a theoretical description.

Earlier FMD calculations \cite{fmd04,chernykh07,chernykh10} as well as AMD calculations \cite{enyo07} investigated both the ground state band and the cluster states but treated states in the continuum in a bound-state approximation. This might be justified for the very narrow Hoyle state at 7.65~MeV but is certainly very unreliable for the broad resonances observed higher up in energy.

\begin{figure}[b]
	\centerline{
	\includegraphics[width=0.90\textwidth]{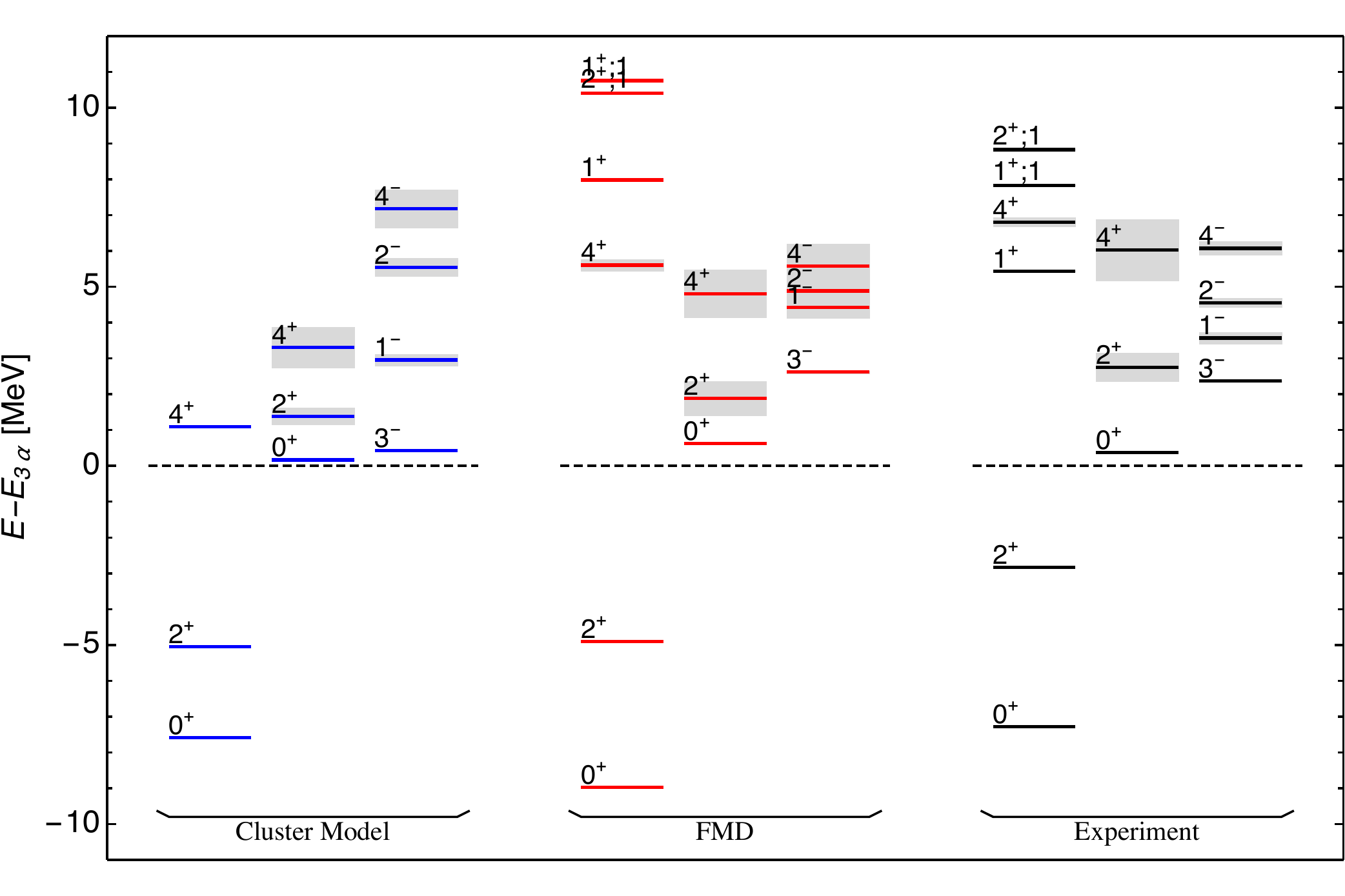}}
	\caption{$^{12}$C energy spectra obtained with a microscopic cluster model (left) and FMD (middle) compared with experiment (right). Energies are given with respect to the $3-\alpha$ threshold. Resonance widths are indicated by shaded areas.}
	\label{fig:c12}
\end{figure}

To address these questions we extended our approach with a proper treatment of the continuum in a similar way as explained in the last section. We first performed a study within the microscopic $\alpha$-cluster model with full antisymmetrization and a phenomenological two-body interaction \cite{neff14} similar to earlier calculations by Descouvemont and Baye\cite{descouvemont87} and Arai\cite{arai06}. In this cluster model the internal region the Hilbert space is built from three-$\alpha$ configurations on a triangular grid without any restrictions. In the external region $^8$Be+$^4$He configurations are added. In principle we have to deal with a real three-body continuum. However, in $^8$Be the Coulomb repulsion of the two $\alpha$ particles is compensated by the nuclear attraction which lowers the energy surface of the $^8$Be+$^4$He configurations compared to free $\alpha$ particles significantly. It turns out that it is important to include not only the $^8$Be ground state but also the excited $2^+$ state. The inclusion of additional $^8$Be states changes the final results however only slightly.

The microscopic $R$-matrix method\cite{descouvemont10} is used to match the microscopic wave functions in the internal region to the asymptotic behavior described by point-like $^8$Be- and $\alpha$-clusters. For bound states the asymptotics is given by Whittaker functions, for resonances we match to purely outgoing Coulomb wave functions. The energies of these Gamow states are complex, with the real part giving the resonance position and the imaginary part the resonance width. We can also obtain scattering states with real energies by matching to linear combinations of incoming and outgoing Coulomb wave functions which provides the full scattering matrix. For the lower lying resonances shown in Fig.~\ref{fig:c12} the resonances are well separated and resonance parameters obtained from the Gamow states and from the phase shifts are consistent. If one goes higher in energy the situation becomes more complicated and it is mostly impossible to isolate individual resonances. It is however possible to calculate transition strengths as a function of energy.

\begin{figure}[b]
	\centerline{
	\includegraphics[width=0.95\textwidth]{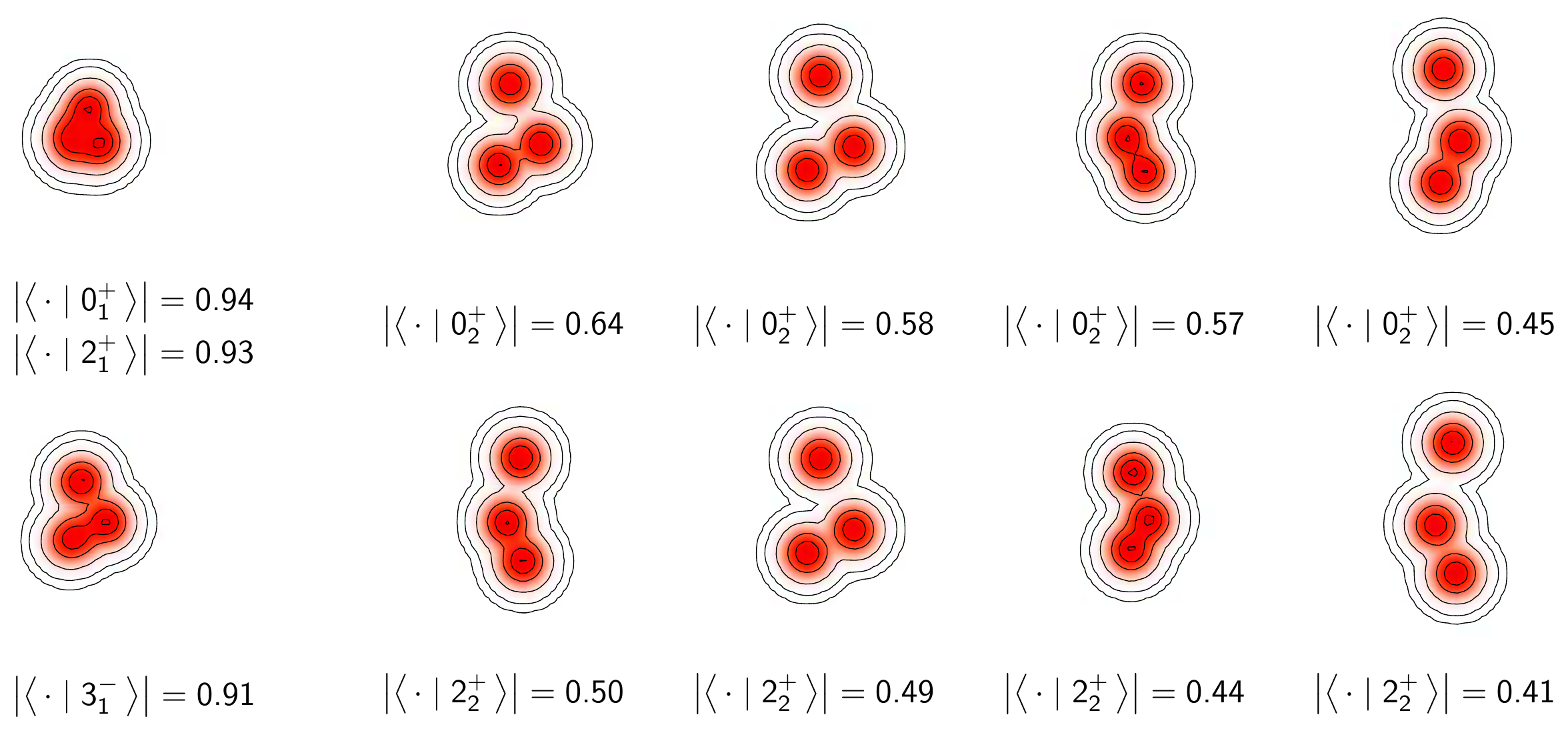}}
	\caption{(Left) Intrinsic densities of dominant FMD basis states and the amplitudes of these basis states in the $0_1^+$, $2_1^+$ and $3_1^-$ eigenstates. (Right) Intrinsic densities of FMD basis states that contribute with large amplitudes to the strongly clustered $0_2^+$ and $2_2^+$ states. Note that the basis states are not orthogonal.}
	\label{fig:c12-intrinsic}
\end{figure}

Recently we also added the continuum to the full FMD calculation. Here basis states in the internal region are obtained by using variation after projection on angular momentum and parity. For each spin we first vary the parameters of the many-body state to obtain the lowest possible energy. A second basis state is then obtained by minimizing the energy of the second state with respect to its parameters keeping the first state fixed. We further increase the model space by using the radii of the intrinsic states as generator coordinates. In Fig.~\ref{fig:c12-intrinsic} the intrinsic densities for some basis states that have a large overlap with the eigenstates are shown. A triangular structure with strongly overlapping $\alpha$-clusters (that also has a large overlap with a $p$-shell shell model wave function) can already be seen in the ground state band members $0_1^+$ and $2_1^+$. An underlying $\alpha$-structure is also visible in the $3_1^-$ state but with a larger spatial extension. For the $0_2^+$ Hoyle state and the $2_2^+$ state the spatial extension is even larger and the $^8$Be+$\alpha$ structure is reflected in the large overlap with several configurations with more or less open triangles.

The $^8$Be clusters that are needed for the description in the external region are obtained by diagonalization in a basis of FMD many-body states and of $\alpha$-$\alpha$ configurations, treating them as pseudostates. We include two $0^+$ states, two $2^+$ states and a $4^+$ state for $^8$Be. 

In Fig.~\ref{fig:c12} we compare the spectra containing bound states and resonances obtained with the microscopic cluster model and with FMD. The cluster model can not describe spin-flip states like the $1^+$ states or the $2^+$ ($T=1$) state. The FMD calculations show in general a good agreement with experimental observations. For example we obtain the $4^+$ from the ground state band, the $4^+$ state of the Hoyle state band and the $4^-$ at roughly the same energy, in good agreement with experiment. Future studies will focus on transitions into the continuum. Apart from the investigations of the monopole strength\cite{chernykh07}, recent experiments studied transitions into the second $2^+$ state and the $1^-$ state \cite{zimmerman13}. Also Gamow-Teller transitions from $^{12}$B and $^{12}$N and electromagnetic transitions from the $2^+$ ($T=1$) state into the $^{12}$C continuum have been measured carefully\cite{hyldegaard10}.

\bibliographystyle{ws-rv-van-simple}
\bibliography{fmd}

\end{document}